\documentclass[aps,prb,reprint,groupedaddress,showpacs,floatfix]{revtex4-1}

\usepackage{graphicx}
\usepackage{dcolumn}
\usepackage{amsmath}

\begin{document}

\title{Exchange-correlation potentials for inhomogeneous electron systems in two dimensions from exact diagonalization: comparison with the local-spin-density approximation}

\author{Ilja Makkonen}
\email[]{ilja.makkonen@aalto.fi}
\author{Mikko M. Ervasti}
\author{Ville Kauppila}
\author{Ari Harju}
\affiliation{Department of Applied Physics and Helsinki Institute of Physics, Aalto University, P.O. Box 14100, FI-00076 AALTO, Espoo, Finland}

\date{\today}

\begin{abstract}
We consider electronic exchange and correlation effects in density-functional calculations of two-dimensional systems. Starting from wave function calculations of total energies and electron densities of inhomogeneous model systems, we derive corresponding exchange-correlation potentials and energies. We compare these with predictions of the local-spin-density approximation and discuss its accuracy. Our data will be useful as reference data in testing, comparing and parametrizing exchange and correlation functionals for two-dimensional electronic systems.
\end{abstract}


\maketitle

\section{Introduction}

Many modern nanoelectronic devices such as quantum dots rely on reduced dimensionality. Two-dimensional (2D) electron systems can be well described using a 2D Hamiltonian in the effective mass approximation.~\cite{Reimann02,Saarikoski10} The density-functional theory (DFT) can be applied to describe electronic properties in 2D as well as in three dimensions (3D). However, 3D functionals for electronic exchange and correlation are not, in general, as such applicable for systems with reduced dimensionality.~\cite{Kim00,Pollack00,Constantin08a,Constantin08b}

For electronic systems confined in 2D there exists, in addition to two local-spin-density approximation (LSDA) parametrizations,~\cite{Tanatar89,Attaccalite02} more recent approximations such as local functionals,~\cite{Pittalis08,Sakiroglu10} a Thomas-Fermi-type explicit functional of the density,~\cite{Pittalis09c} generalized gradient approximations (GGAs),~\cite{Pittalis09b,Pittalis10} and several Laplacian-level functionals (meta-GGAs),~\cite{Pittalis07,Pittalis09,Pittalis09d,Rasanen10,Pittalis10c,Pittalis10b} describing either the exchange or correlation energy/potential or both. Also the optimized effective potential method~\cite{Talman76} and orbital functionals have been applied in 2D.~\cite{Helbig08} However, these approximations and their possible combinations remain relatively untested so far, and their predictions have not been extensively compared against one another or against same accurate reference data.

In many cases, approximations for the exchange and correlation energy functional are derived assuming that the electron density is slowly varying and by starting from many-body calculations made for homogeneous electron gas (for 2D examples, see for example Refs.~\onlinecite{Tanatar89,Attaccalite02}) and the local-density or local-spin-density approximation (LDA/LSDA). An alternative route taken in this work is to consider inhomogeneous model systems. Starting from accurate densities and total energies, one can obtain accurate exchange and correlation potentials and energies for the model systems. Given an accurate ground-state density of a model system, the Kohn-Sham potential and the exchange-correlation potential follow uniquely by virtue of the Hohenberg-Kohn theorem. Since the exact exchange-correlation functional is unknown, the potential can only be obtained by iterative potential inversion techniques, which can provide the potential reproducing a given accurate density. Approximate exchange-correlation functional should produce for the same density the same exchange and correlation energies and potentials so that both the energy and the electron density would converge to their correct values.

The main focus of this Article is on generating accurate reference data for testing, comparing, and creating new functionals and their parametrizations for the electronic exchange and correlation in 2D. In order to be able to access the exchange-correlation potentials and energies of our model systems, we calculate highly converged electron densities using the exact diagonalization technique. The reference data can be used to benchmark approximate exchange-correlation functionals. Most importantly, this can include comparing approximate exchange and correlation potentials, whose accuracy determines the accuracy of the electron density in self-consistent DFT calculations, with our accurate ones. To give an idea of the accuracy of present functionals, we study the accuracy of the LSDA parametrization by Attaccalite \emph{et al}.~\cite{Attaccalite02} The LSDA is a unique functional, which becomes accurate in the limit of slow density variations, and many approximations build on top of it. As such, it is the most important starting point of comparison.

The Article is structured as follows. In Sec.~\ref{formalism}, we review briefly the formalism related to DFT calculations within the Kohn-Sham method. Section~\ref{computational} presents our model systems and the specifics of our many-body and DFT calculations. Section~\ref{results} contains our results and discussion, and Sec.~\ref{conclusions} presents our conclusions.

\section{Formalism}\label{formalism}

In the spin-polarized version of the Kohn-Sham method for the density-functional theory, the energy functional is written in terms of the two spin densities $n_{\uparrow}(\mathbf{r})$ and $n_{\downarrow}(\mathbf{r})$ and the total density
$n(\mathbf{r})=n_{\uparrow}(\mathbf{r})+n_{\downarrow}(\mathbf{r})$,
\begin{align}\label{KSfunctional}
 E_{\text{KS}}[n_{\uparrow},n_{\downarrow}] & =-\frac{1}{2}\sum_{\sigma=\uparrow,\downarrow}\sum_{\text{occ}}\int d\mathbf{r}\,\psi_{i}^{\sigma}(\mathbf{r})\nabla^{2}\psi_{i}^{\sigma}(\mathbf{r})\nonumber\\
& + \frac{1}{2}\int\int d\mathbf{r}\,d\mathbf{r}'\,\frac{n(\mathbf{r})n(\mathbf{r}')}{|\mathbf{r}-\mathbf{r}'|}\nonumber\\
 & + \int d\mathbf{r}\,v_{\text{ext}}(\mathbf{r})n(\mathbf{r})+E_{\text{xc}}[n_{\uparrow},n_{\downarrow}],
\end{align}
where the summation in the first term is over occupied Kohn-Sham orbitals. Here we use the Hartree atomic units. Above, the first term is the kinetic energy of an auxiliary non-interacting system with spin densities equal to those of the interacting one, the second one the classical Hartree energy, the third one the interaction energy of the electron system with the external potential, and the last term the exchange-correlation energy, which needs to be approximated. The orbitals $\psi_{i}^{\sigma}(\mathbf{r})$ of the auxiliary non-interacting system are solved from the single-particle equations
\begin{equation}\label{KS}
 \left[-\frac{1}{2}\nabla^{2}+v_{\text{eff}}^{\sigma}(\mathbf{r})\right]\psi_{i}^{\sigma}(\mathbf{r})=\varepsilon_{i}^{\sigma}\psi_{i}^{\sigma}(\mathbf{r}),
\end{equation}
in which the effective potentials can be written as
\begin{equation}\label{veff}
 v_{\text{eff}}^{\sigma}(\mathbf{r})=\int d\mathbf{r}'\,\frac{n(\mathbf{r}')}{|\mathbf{r}-\mathbf{r}'|}+v_{\text{ext}}(\mathbf{r})+v_{\text{xc}}^{\sigma}(\mathbf{r}),
\end{equation}
where $v_{\text{ext}}(\mathbf{r})$ is the external potential and $v_{\text{xc}}^{\sigma}(\mathbf{r})$ the exchange-correlation potential, a functional derivative of the exchange-correlation energy,
\begin{equation}
 v_{\text{xc}}^{\sigma}(\mathbf{r})=\frac{\delta E_{\text{xc}}[n_{\uparrow},n_{\downarrow}]}{\delta n_{\sigma}(\mathbf{r})}.
\end{equation}
The densities expressed in terms of orbitals are
\begin{equation}\label{density}
 n_{\sigma}(\mathbf{r})=\sum_{\text{occ}}|\psi_{i}^{\sigma}(\mathbf{r})|^{2},
\end{equation}
where one again sums over occupied orbitals. Equations~(\ref{KS})--(\ref{density}) are iterated self-consistently until the effective potentials $v_{\text{eff}}^{\sigma}(\mathbf{r})$ and densities $n_{\sigma}(\mathbf{r})$ are consistent with one another.

The Hohenberg-Kohn theorem applied to the non-interacting Kohn-Sham system guarantees that if the exact ground-state densities $n_{\sigma}(\mathbf{r})$ are known, the effective potentials $v_{\text{eff}}^{\sigma}(\mathbf{r})$ follow uniquely, as long as the densities are noninteracting $v$ representable. Several numerical algorithms~\cite{Wang93,Umrigar94,Gorling92,vanLeeuwen94,Gritsenko95} exist for finding the effective potentials corresponding to given ground-state densities.

In this work, we construct numerically accurate exchange-correlation potentials for specific model systems in two dimensions according to the above prescription, by starting from accurate wave function calculations yielding the electron densities. We compare the obtained potentials to ones predicted by the LSDA parametrization by Attaccalite \emph{et al.}~\cite{Attaccalite02} and discuss their differences. We also obtain information on total energies, exchange and correlation energies, and how these are described by the LSDA. The data derived from accurate wave function calculations can also serve as a database useful for comparing and benchmarking present and future functionals.

\section{Computations and model systems}\label{computational}

In order to be able to analyze the behavior of numerically accurate exchange-correlation potentials of inhomogeneous electron systems in 2D and compare them with ones predicted by the LSDA, we consider a few specific model systems and calculate their total energies and electron densities using the exact diagonalization technique (ED).~\cite{Gylfadottir06}
To reduce the potential inversion procedure to a one-dimensional problem, we restrict ourselves to systems that are radially symmetric. We use either a harmonic potential ($v_{\text{ext}}(r)=\omega^{2}r^{2}/2$) or a circular hard-wall potential (infinite potential beyond some given radius $R$ and zero within) to confine the electrons. Using methods described below, we invert the effective potentials $v_{\text{eff}}^{\sigma}(\mathbf{r})$ reproducing the spin densities $n_{\sigma}(\mathbf{r})$ of each given model. 

In the special case of two electrons in a spin singlet, the effective potential can be obtained directly from the density.~\cite{Filippi94} The Kohn-Sham orbital is calculated as [see Eq.~(\ref{density})]
\begin{equation}
\psi^{\sigma}(\mathbf{r})=\sqrt{n_{\sigma}(\mathbf{r})},
\end{equation}
and the effective and exchange-correlation potentials can then be solved from Eqs.~(\ref{KS}) and (\ref{veff}).

In a more general case, we use an iterative potential inversion algorithm.~\cite{vanLeeuwen94,Gritsenko95} In the course of the iteration, a new approximation for the effective potential  [$v_{\text{eff}}^{\sigma,i+1}(\mathbf{r})$] is calculated from previous one [$v_{\text{eff}}^{\sigma,i}(\mathbf{r})$], and the corresponding density $n_{\sigma}^{i}(\mathbf{r})$ as
\begin{equation}\label{iteration}
v_{\text{eff}}^{\sigma,i+1}(\mathbf{r})=\frac{n_{\sigma}^{i}(\mathbf{r})+a}{n_{\sigma}(\mathbf{r})+a}v_{\text{eff}}^{\sigma,i}(\mathbf{r}),
\end{equation}
where $n_{\sigma}(\mathbf{r})$ is the reference (ED) density whose generating exchange-correlation potential we want to determine, and $a>0$ is a smoothing parameter removing the effect of density tails. In order to keep the iteration stable, we start with a large $a$ and decrease its value as the potential starts to converge. Also, we 
take care that the prefactor of Eq.~(\ref{iteration})
does not deviate too much from unity.~\cite{vanLeeuwen94} The behavior of the algorithm has been found to depend on the zero level of the potential.~\cite{vanLeeuwen94} An empirical modification of the scheme and Eq.~(\ref{iteration}) we have found to work well for harmonically confined systems is to align $v_{\text{eff}}^{\sigma}(\mathbf{r})$ to be negative and to use an inverse prefactor corresponding to iteration to the opposite direction. Then the potential is raised where the density is too high and vice versa.

As we vary the parameters of the confining potential and the number of electrons in the system, we get a corresponding set of densities and effective potentials to analyze. We can, for instance, compare the exchange-correlation potentials and energies determined by Eqs.~(\ref{veff}) and~(\ref{KSfunctional}) to results predicted by the LSDA.

In the case of the harmonic confinement with varying $\omega$, we focus on two electrons in a spin singlet, 4 electrons with $(L,S)=(0,1)$, and 6 with $(L,S)=(0,0)$, and in the case of the hard-wall confinement with varying $R$, on two electrons in a spin singlet state. For the harmonically confined systems we either use expansion in relative coordinates (two-electron case, number of terms taken high enough to give numerically exact results) or the simple-harmonic-oscillator basis (4 to 6 electrons) and full ED with up to 55 single-particle basis functions. For the hard-wall systems, we use Bessel functions and do full ED with up to 50 single-particle basis functions. In our DFT calculations we use a Bessel function basis.

The Hamiltonian of interacting Coulomb particles in an external potential $v_{\text{ext}}(\mathbf{r})$,
\begin{equation}\label{aunits}
 H=\sum_{i}\left[-\frac{1}{2}\nabla_{i}^{2}+v_{\text{ext}}(\mathbf{r}_{i})\right]%\nonumber\\
+\frac{1}{2}\sum_{i\ne j}\frac{1}{|\mathbf{r}_{i}-\mathbf{r}_{j}|},
\end{equation}
can be in many interesting examples of confining potential expressed via nondimensionalization effectively as
\begin{equation}\label{nunits}
\frac{H}{\gamma^{2}}=\sum_{i}\left[-\frac{1}{2}\nabla_{i}^{2}+v'_{\text{ext}}(\mathbf{r}_{i})\right]%\nonumber\\
+\frac{1}{2}\frac{1}{\gamma}\sum_{i\ne j}\frac{1}{|\mathbf{r}_{i}-\mathbf{r}_{j}|},
\end{equation}
where $v'_{\text{ext}}$ is now independent of the parameter characterizing the confinement whose effect now is to determine a unit system, namely natural length and energy scales for the specific model potential, and scale the strength of the electron-electron interaction. The nondimensionalized Eq.~(\ref{nunits}) is obtained from Eq.~(\ref{aunits}) by substituting $\mathbf{r}\rightarrow\mathbf{r}/\gamma$, choosing $\gamma^{2}$ as the energy unit and the value of $\gamma$ so that $\mathbf{r}$ becomes a dimensionless variable, and identifying the new $v'_{\text{ext}}$ independent of the confinement parameter. In Eq.~(\ref{nunits}), the natural units of energy are $\gamma^{2}$ and those of the length $1/\gamma$, and the interaction is scaled by $1/\gamma$. For instance, for the harmonic confining potential $v_{\text{ext}}(r)=\omega^{2}r^{2}/2$, $\gamma=\sqrt{\omega}$ and $v'_{\text{ext}}(r)=r^{2}/2$, and the units for energy and length are $\omega$ and $1/\sqrt{\omega}$ (HO units). Then in the HO unit system, having a confinement of $\omega$ corresponds simply to scaling the interaction by $1/\sqrt{\omega}$ while keeping the external potential fixed ($\omega\equiv 1$). For the hard-wall potential with confinement radius $R$, $\gamma=1/R$ and the units for energy and length are $1/R^{2}$ and $R$ (HW units).

\section{Results and discussion}\label{results}

\subsection{Energies}

Our model systems are summarized in Table~\ref{Models}, where we also list our ED total energies and DFT total energies calculated using the LSDA parametrization by Attaccalite \emph{et al}.~\cite{Attaccalite02} All the LSDA results in this article correspond to ground-state densities of self-consistent LSDA calculations. Using ED densities and corresponding orbitals would not affect our conclusions. The potential inversion procedure provides us, in addition to the accurate effective potential, also with accurate Kohn-Sham orbitals. Using these and the Kohn-Sham energy functional [Eq.~(\ref{KSfunctional})] we can calculate the exact ground-state exchange-correlation energy, $E_{xc}$, for any given model system. These, along with the corresponding approximate LSDA values, are also listed in Table~\ref{Models}. In the case of the spin-singlet two-electron systems the exchange energy is simply~\cite{Filippi94}% the self-interaction correction,
\begin{equation}
 E_{x}[n_{\uparrow},n_{\downarrow}]=-\frac{1}{4}\int\int d\mathbf{r}\,d\mathbf{r}'\,\frac{n(\mathbf{r})n(\mathbf{r}')}{|\mathbf{r}-\mathbf{r}'|},
\end{equation}
i.e., minus one half times the Hartree energy. This allows us to easily decompose the exchange-correlation energies of these systems into exchange and correlation parts.
%\begin{turnpage}
\begingroup
\squeezetable
\begin{table*}
\caption{Summary of our model systems and calculated energies. The table includes the systems' descriptions, shape of the confining potential and the related parameter $R$ or $\omega$, our ED total energies, $E_{\text{tot}}$, the LSDA total energies $E_{\text{tot}}^{\text{LSDA}}$, calculated using the parametrization of Ref.~\onlinecite{Attaccalite02}, the exchange-correlation energies inverted from the ED calculation, $E_{\text{xc}}$, and the corresponding LSDA values, $E_{\text{xc}}^{\text{LSDA}}$. For two-electron systems we also show their decompositions into exchange and correlation energies, $E_{\text{x}}$ and $E_{\text{c}}$. Natural units determined by the external potential are used throughout.\label{Models}}
\begin{ruledtabular}
\begin{tabular}{lddddddddd}
System & R / \omega & E_{\text{tot}} & E_{\text{tot}}^{\text{LSDA}} & E_{\text{xc}} & E_{\text{xc}}^{\text{LSDA}} & E_{\text{x}} & E_{\text{x}}^{\text{LSDA}} & E_{\text{c}} & E_{\text{c}}^{\text{LSDA}}\\
\hline
hard-wall&1&8.1160&8.2773&-2.6993&-2.5315&-2.4882&-2.2700&-0.2111&-0.2614\\
$N=2$&2&10.0483&10.3287&-5.5439&-5.2483&-4.8055&-4.4042&-0.7384&-0.8441\\
$(L,S)=(0,0)$&3&11.7267&12.1076&-8.4711&-8.0587&-7.0150&-6.4553&-1.4561&-1.6035\\
&4&13.2424&13.6996&-11.4419&-10.9241&-9.1563&-8.4526&-2.2856&-2.4716\\
&5&14.6492&15.1541&-14.4324&-13.8252&-11.2500&-10.4135&-3.1824&-3.4118\\
&6&15.9786&16.5019&-17.4288&-16.7520&-13.3072&-12.3490&-4.1216&-4.4031\\
&7&17.2502&17.7631&-20.4236&-19.6997&-15.3343&-14.2674&-5.0893&-5.4324\\
&8&18.4764&18.9511&-23.4133&-22.6668&-17.3357&-16.1749&-6.0775&-6.4920\\
&9&19.6655&20.0758&-26.3966&-25.6536&-19.3145&-18.0769&-7.0821&-7.5769\\
&10&20.8232&21.1440&-29.3738&-28.6623&-21.2739&-19.9785&-8.0999&-8.6839\\
\hline
harmonic& 0.1 & 4.40792 & 4.49851 &  -3.76280 & -3.61541 & -2.88852 & -2.59544 & -0.87428
& -1.01997\\
$N=2$ & 0.25 & 3.72056 & 3.81038 & -2.44389 & -2.32795 & -1.97158 & -1.77172 & -0.47231 & -0.55623\\
$(L,S)=(0,0)$ & 0.5 & 3.31954 & 3.39844 & -1.75162 & -1.65694 & -1.46774 & -1.31890 & -0.28388 & -0.33804\\
& 1 & 3.00000 &  3.06553 & -1.24909 & -1.17374 & -1.08639 & -0.97497 & -0.16269 & -0.19876\\
\hline
harmonic & 0.25 & 13.6187 & 13.6974 & -5.3217 & -5.2082 & & & & \\
$N=4$ & 0.5 & 11.7426 & 11.8108 & -3.8449 & -3.7617 & & & & \\
$(L,S)=(0,1)$ & 1 & 10.2807 & 10.3394 & -2.7701 & -2.7034 & & & & \\
\hline
harmonic&0.25&27.961&28.049&-8.147&-8.045\\
$N=6$&0.5&23.610&23.679&-5.897&-5.825\\
$(L,S)=(0,0)$&1&20.198&20.254&-4.257&-4.200\\
\end{tabular}
\end{ruledtabular}
\end{table*}
\endgroup
%\end{turnpage}

Some ED reference results exist for the total energies of harmonically confined systems in the literature. For the two-electron systems our energies are lower than results of Ref.~\onlinecite{Ciftja05}, which are 3.72143 ($\omega=0.25$) and 3.00097 ($\omega=1$). For $\omega=1$ there exists an analytic solution with an energy of 3 (Ref.~\onlinecite{Taut94}). For the 4-electron system with $\omega=0.25$ our energy is only slightly higher than one calculated by Mikhailov,~\cite{Mikhailov02} who obtained 13.6180 using a larger one-body basis. For 6 electrons with $(L,S)=(0,0)$ and $\omega=0.25$ Rontani~\emph{et al.}~\cite{Rontani06} have obtained the energy of 27.98 using 36 single-particle states, which is higher than our result calculated with 55 states.

The ED and LSDA total energies are rather consistent for all systems, the latter one being curiously consistently higher, despite the fact that the DFT total energy is not guaranteed to be variational once the exchange-correlation energy functional is approximated. Similarly, the LSDA exchange-correlation energies are higher, the exchange component being clearly underestimated. %overestimated.
The lower correlation energy predicted by the LSDA partly compensates this. This finding of error cancellation is consistent with the results for He isoelectronic series in 3D.~\cite{Umrigar94}
The mechanisms behind the underestimation of the exchange energy and cancellation of errors between exchange and correlation energies are the same for the 2D as for the 3D LSDA. The underestimation of exchange energy is due to self-interaction, whereas the compensating effect of the overestimation of the correlation energy arises from the exchange-correlation hole sum rule. Since the 2D LSDA is based on a physical system, the 2D uniform electron gas, the sum rule is fulfilled and the integrated errors of the exchange and correlation holes cancel.~\cite{Gunnarson79,Hood98}

\begin{figure}
\includegraphics[width=0.8\columnwidth]{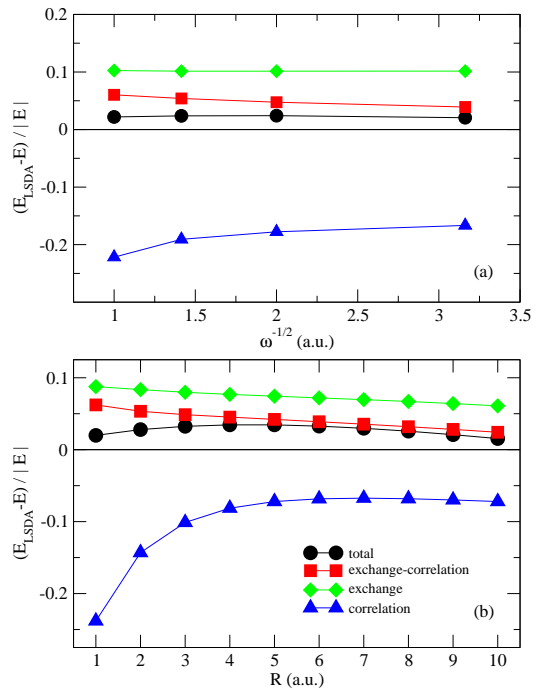}%
\caption{(Color online) Relative errors in the LSDA energy terms (total, exchange-correlation, exchange, and correlation energies) for (a) two-electron systems in harmonic confinement and (b) two electrons in a hard-wall confinement as a function of the characteristic length scale of the system, $1/\sqrt{\omega}$ or $R$.\label{e_vs_width}}
\end{figure}

Figure~\ref{e_vs_width} shows a graphical representation of LSDA's relative errors in the total energy and the exchange-correlation, exchange and correlation energies for the two-electron systems in harmonic and hard-wall confinements as a function of the characteristic length scale of the system, $1/\sqrt{\omega}$ or $R$. The relative errors for the total and exchange energies are rather constant as a function of the systems' length scale. The same applies to the exchange-correlation energy, which consists mostly of the exchange one. The relative error in the correlation energy is larger at stronger confinements (large $\omega$ / small $R$) and becomes smaller at weaker confinement. This does not affect the overall picture much since the correlation energy is a small fraction of the exchange-correlation one, especially at the weakly interacting (large $\omega$ / small $R$) limit. The strongly confined systems are less uniform than their counterparts in weaker confinement. Therefore, the accuracy of the LSDA is worse in this limit.
The accuracy of the LSDA energies is improved with increasing particle number. For two electrons, the level of accuracy is the same for both the hard-wall and harmonic potentials. For 4 electrons, the relative error in the total energy is at most 0.6\% and in the exchange-correlation energy 2.4\%, and for 6 electrons 0.3\% and 1.3\%, respectively.

\subsection{Densities and potentials}

\begin{figure*}
\includegraphics[height=0.74\textheight]{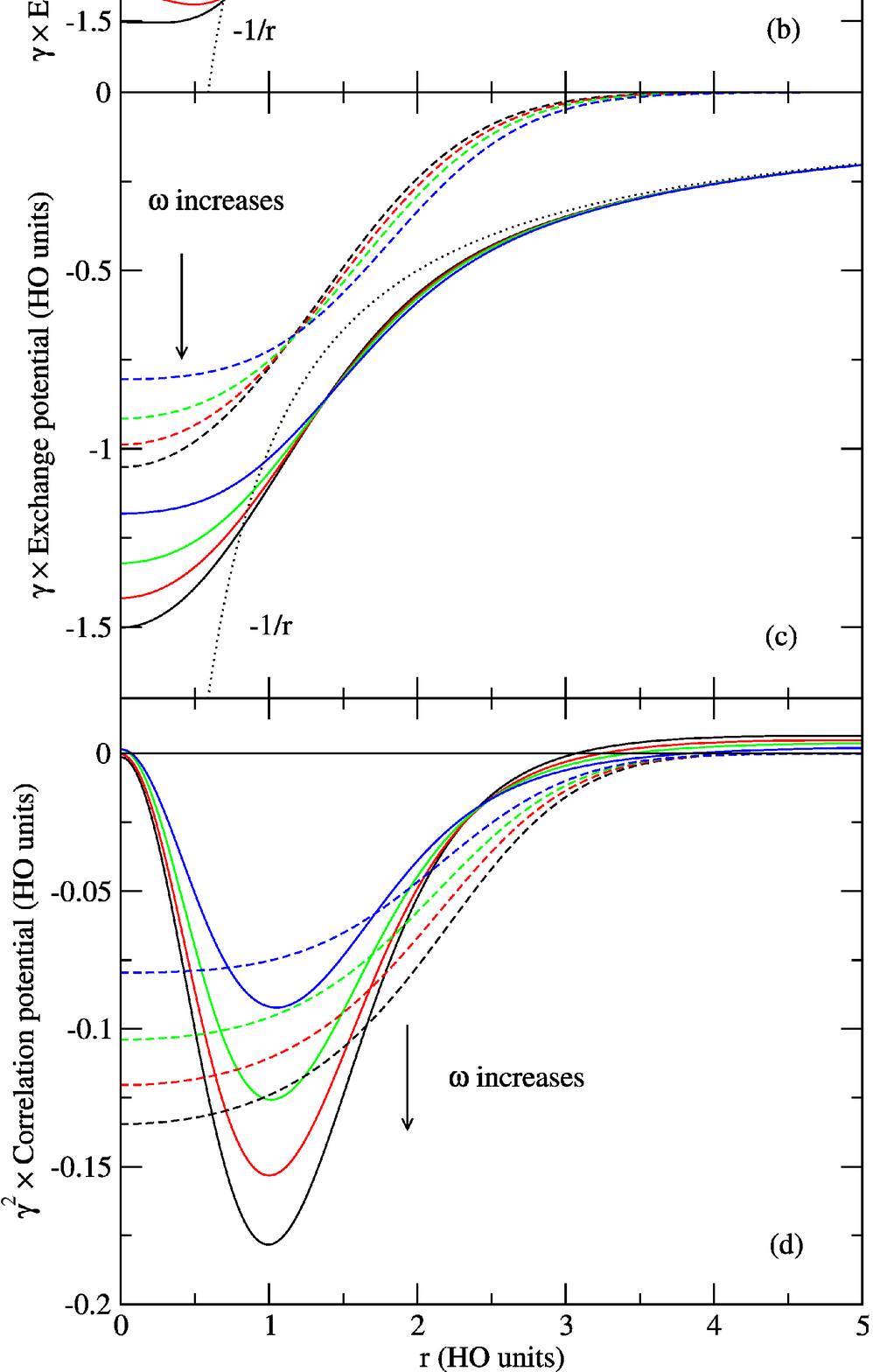}
\includegraphics[height=0.74\textheight]{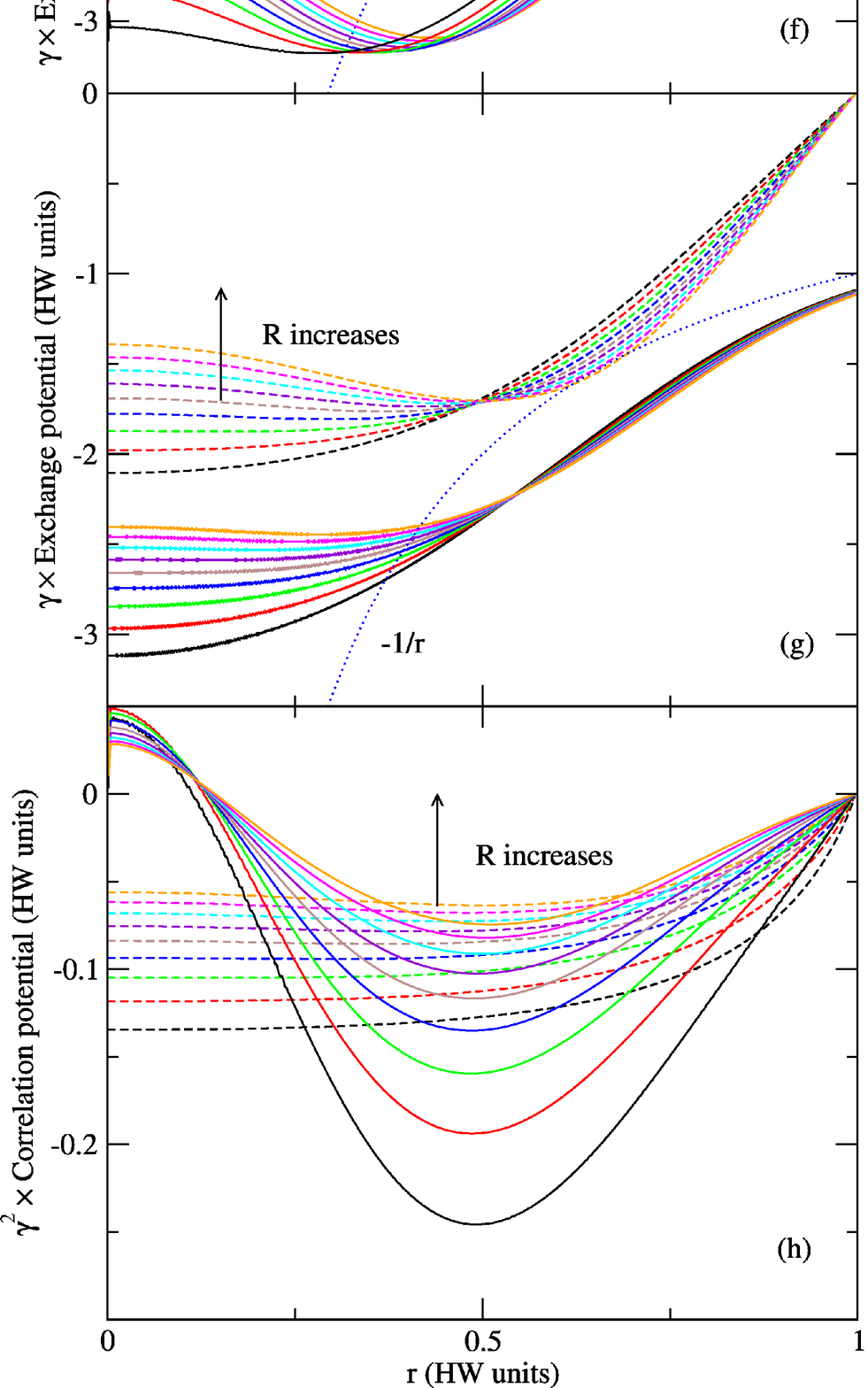}
\caption{(Color online) Densities (a), exchange-correlation potentials (b), exchange potentials (c), and correlation potentials (d) as a function of the distance $r$ for two electrons in a spin singlet in harmonic confinement $\omega$, and (e)--(h) the same quantities for two-electrons in the hard-wall confinement with radius $R$. For the latter the results for $R=2,\ldots,10$ are shown. Solid lines are the accurate results and dash lines the LSDA ones. Thin dotted lines in (b), (c), (f), and (g) show the exact asymptotic $-1/r$ behavior of the exchange potential. Natural units determined by the external potential are used throughout.\label{2e_harm_v_xc}}
\end{figure*}

We begin the comparison of exchange-correlation potentials from the two-electron systems in spin singlet. Figure~\ref{2e_harm_v_xc} shows electron densities and different potential terms (exchange and correlation, exchange, correlation) for the harmonically confined electrons with varying $\omega$ and hard-wall systems with varying $R$.
The scaling we use when representing the data is described and motivated below. For the two-electron systems in spin-singlet, the exchange potential is~\cite{Filippi94}
\begin{equation}
v_{\text{x}}(\mathbf{r})=-\frac{1}{2}\int d\mathbf{r}'\,\frac{n(\mathbf{r}')}{|\mathbf{r}-\mathbf{r}'|}.
\end{equation}

When visualizing the potential terms we use natural units and scale the exchange potential by $\gamma$ and the correlation one by $\gamma^{2}$. Since the exchange-correlation potential consists mostly of the exchange one, we scale it similarly. These scalings provide energy scales at which the potentials are of comparable magnitude and their features can be easily compared. This can be explained using the following scaling relations for the exchange energy,~\cite{Levy85}
\begin{equation}\label{exscaling}
 E_{x}[n_{\lambda}]=\lambda E_{x}[n],
\end{equation}
and the correlation energy,~\cite{Levy87}
\begin{equation}\label{ecscaling}
 E_{c}[n_{\lambda}]=\lambda^{2} E_{c}^{1/\lambda}[n].
\end{equation}
Above, $\lambda$ is an arbitrary scaling parameter not necessarily referring to a transformation between two unit systems, and the scaled density $n_{\lambda}$ is defined as
\begin{equation}
n_{\lambda}(\mathbf{r})=\lambda^{d}n(\lambda\mathbf{r}),
\end{equation}
where $d$ is the dimension (here 2), and $E_{c}^{1/\lambda}[n]$ is the density functional for the correlation energy for a system with density $n$ but with electron-electron interaction scaled by $1/\lambda$. The corresponding scaling relations for the exchange and correlation potentials are analogous. For the exchange potential,\cite{OuYang90}
\begin{equation}\label{vxscaling}
 v_{x}([n_{\lambda}];\mathbf{r})=\lambda v_{x}([n];\lambda\mathbf{r}),
\end{equation}
and for the correlation potential,
\begin{equation}\label{vcscaling}
 v_{c}([n_{\lambda}];\mathbf{r})=\lambda^{2} v_{c}^{1/\lambda}([n];\lambda\mathbf{r}).
\end{equation}
The latter one follows from Eq.~(\ref{ecscaling}) similarly as Eq.~(\ref{vxscaling}) is derived from Eq.~(\ref{exscaling}) (see Ref.~\onlinecite{OuYang90}). We apply Eqs.~(\ref{vxscaling}) and~(\ref{vcscaling}) in such a way that $n$ corresponds to the density of a given system expressed in atomic units and $n_{\lambda}$ the same density scaled to natural units ($\lambda=\gamma$). Obviously the interaction strengths of the functionals in Eqs.~(\ref{vxscaling}) and~(\ref{vcscaling}) do not match with those of the above unit systems, 1 and $1/\gamma$, [see Eqs.~(\ref{aunits}) and~(\ref{nunits})], but, nevertheless, the scaling relations motivate a consistent visual representation.

There is quite a good agreement in the densities, Fig.~\ref{2e_harm_v_xc}(a), between the accurate ED calculations and the LSDA ones for the harmonically confined two-electron systems in the weakly correlated cases (large $\omega$). At smaller $\omega$, the LSDA densities are monotonous while the accurate ones develop a side peak. Also for the electrons in the hard-wall trap [Fig.~\ref{2e_harm_v_xc}(e)], the agreement is best in the weakly correlated limit (small $R$). The trend in the accuracy is here opposite to the one seen for exchange and correlation energies in Fig.~\ref{e_vs_width}, where the energies are in better agreement in the strongly correlated (uniform system) limits. Integrated quantities such as the energy can behave differently from local quantities such as the density or the potential in the sense that integrated quantities can be more accurate due to cancellation of local errors. Especially in the case of the harmonic potential the agreement in the densities is bad only at a small area at the center of the quantum well.

In the exchange-correlation potential of the harmonically confined electrons, Fig.~\ref{2e_harm_v_xc}(b), the most marked differences are the different asymptotic behavior ($-1/r$ vs Gaussian decay) due to the lack of self-interaction correction in the LSDA exchange potential, see also the exchange potential in Fig.~\ref{2e_harm_v_xc}(c), and the resulting vertical shift. For the hard-wall systems, the differing behavior of the potentials close to the wall is even more pronounced due to the rapid decay of the charge density, see Figs.~\ref{2e_harm_v_xc}(f)--(g). The shapes of the exchange-correlation potentials close to the potential well center differ, especially in the case of harmonically confined systems. The unphysical monotonousness of the LSDA exchange-correlation potentials of the harmonically confined systems arises from the LSDA's local character and the monotonic behavior of the LSDA densities. Separate comparisons of the exchange and correlation potentials of these systems, Figs.~\ref{2e_harm_v_xc}(c) and~\ref{2e_harm_v_xc}(d), reveal that the non-monotonicity of the accurate exchange-correlation potential arises from the correlation part. It is also noteworthy that the correlation potential changes sign to positive at large radii and decays to zero from above. This behavior is familiar from 3D systems.~\cite{Umrigar94,Filippi94} In the hard-wall systems, [Figs.~\ref{2e_harm_v_xc}(f)--(h)] the accurate correlation potential is again non-monotonous making the exchange-correlation potential non-monotonous for all values of $R$. On the other hand, the behavior of the accurate exchange potential is, for almost all values of $R$ apart from the largest considered, monotonous, whereas the LSDA exchange potential displays a stronger non-monotonous behavior following the trends of the local electron density. In conclusion, for the hard-wall system the LSDA exchange and correlation potentials take different roles than the accurate ones and the resulting canceling of errors leads to exchange-correlation potentials that are qualitatively correct. This is similar to the error cancellation between exchange and correlation energies discussed above.

\begin{figure*}[t]
\includegraphics[width=0.5\textwidth]{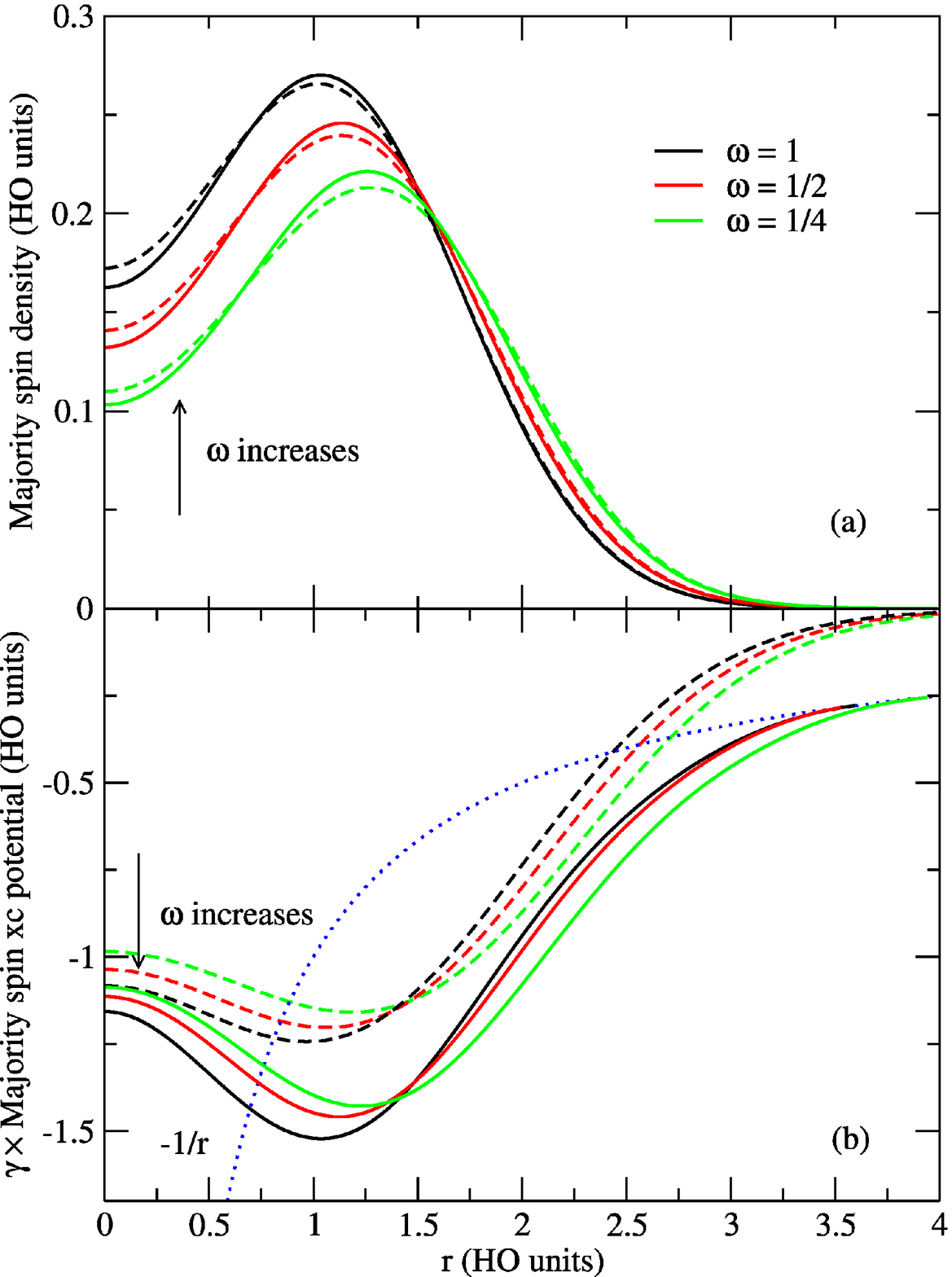}%
\includegraphics[width=0.5\textwidth]{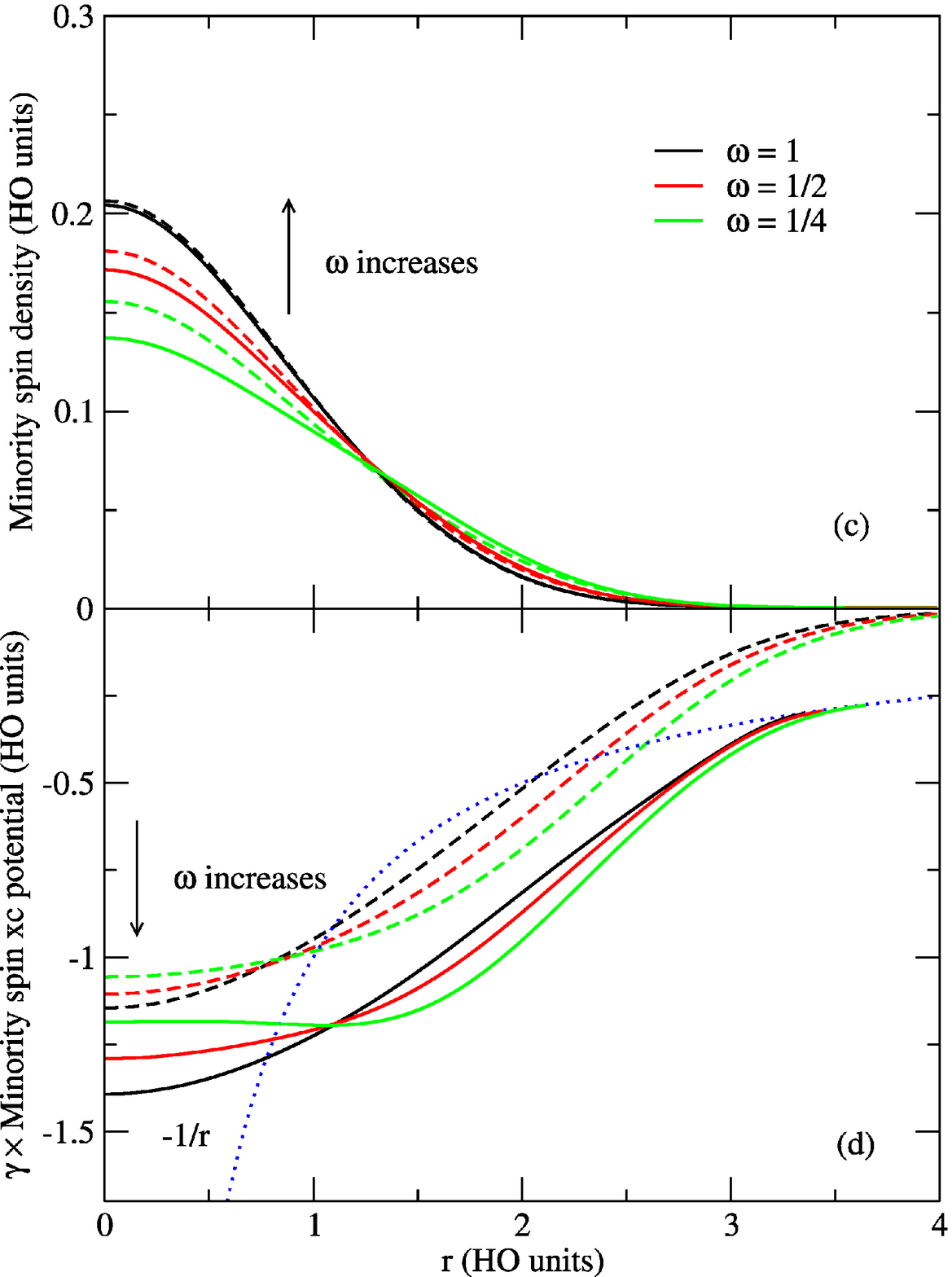}%
\caption{(Color online) The majority spin density (a) and majority spin exchange-correlation potential (b) as a function of the radius for 4 electrons with $(L,S)=(0,1)$ in harmonic confinement $\omega$, and (c)--(d) the same quantities for the minority spin. Solid lines are the accurate results and dash lines the LSDA ones. The thin dotted line shows the exact asymptotic $-1/r$ behavior of the exchange potential. HO units are used.\label{4e_potentials_majority}}
\end{figure*}

The next closed-shell system is 4 electrons with $(L,S)=(0,1)$. Figure~\ref{4e_potentials_majority} shows the spin densities and spin-dependent exchange-correlation potentials for varying $\omega$. For the majority spin [Fig.~\ref{4e_potentials_majority}(a)--(b)], both the densities and exchange-correlation potentials are described rather accurately. For the minority spin [Fig.~\ref{4e_potentials_majority}(c)--(d)], the accuracy decreases with decreasing $\omega$. The LSDA exchange-correlation potentials are, however, qualitatively correct apart their wrong asymptotics and less repulsive shape at the origin.

\begin{figure}[t]
\includegraphics[width=0.8\columnwidth]{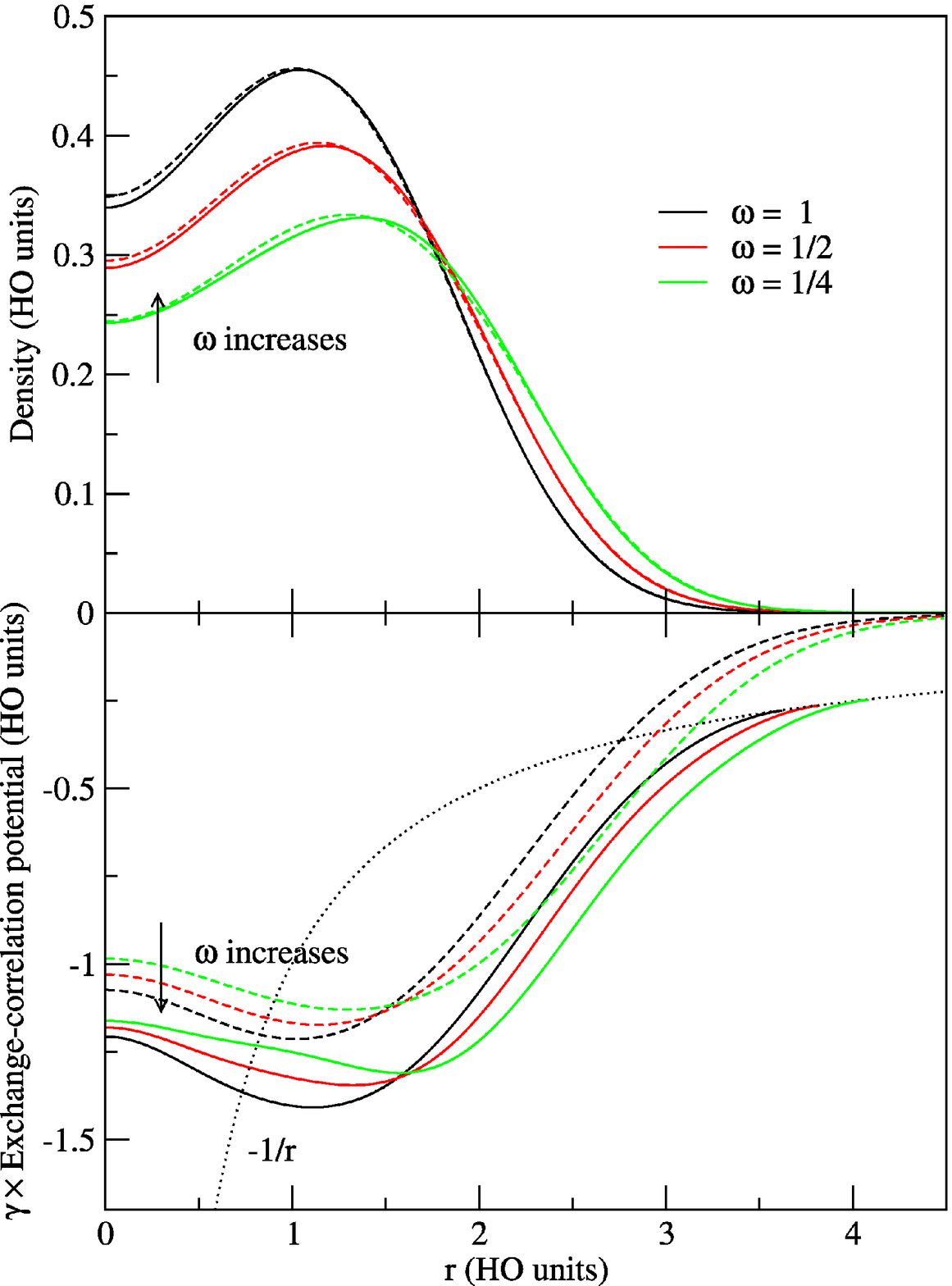}
\caption{(Color online) The electron density (a) and potential (b) as a function of the radius for 6 electrons with $(L,S)=(0,0)$ in harmonic confinement $\omega$. Solid lines are the accurate results and dash lines the LSDA ones. The thin dotted line shows the exact asymptotic $-1/r$ behavior of the exchange potential. HO units are used.\label{6e_potentials}}
\end{figure}

Finally, Fig.~\ref{6e_potentials}(a) shows densities and Fig.~\ref{6e_potentials}(b) exchange-correlation potentials for 6 electrons with $(L,S)=(0,0)$ in harmonic potential with varying $\omega$. For these unpolarized systems with higher average electron densities, the accuracy of the LSDA is remarkable both in the densities and exchange-correlation potentials. Only at small $\omega$ the LSDA and accurate exchange-correlation potentials start to display differences. %Figure~\ref{v_xc_vs_n_6e} shows the exchange-correlation potential as a function of the local electron density. The result is consistent with the 2-electron systems, see Fig.~\ref{v_xc_vs_n}. In this case the potentials are closer to being local functions of the density consistently with the good agreement between the accurate and LSDA exchange-correlation energies.

\begin{figure*}[t]
\includegraphics[width=0.9\textwidth]{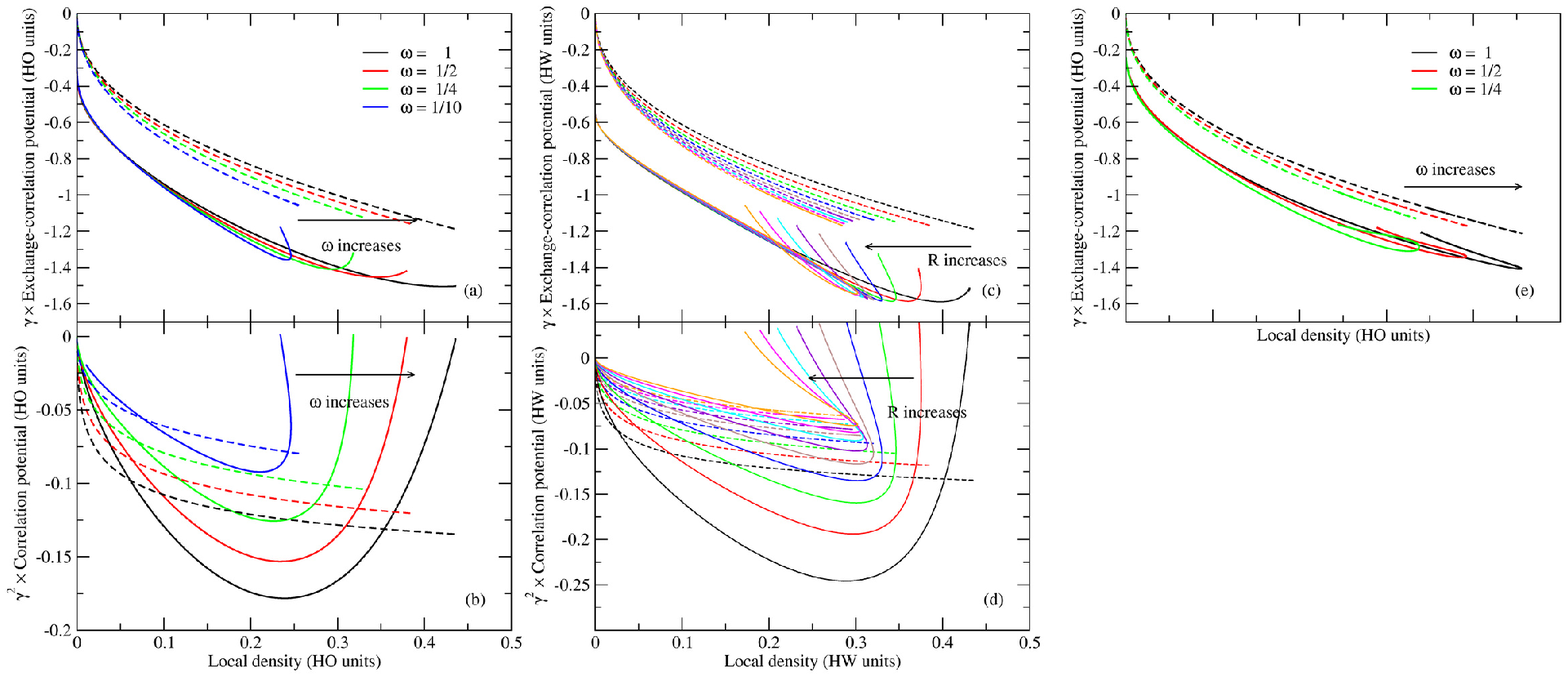}%
\caption{(Color online) The exchange-correlation potential (a) and the correlation potential (b) as a function of the local electron density for two electrons in spin singlet in harmonic confinement $\omega$ [see Fig.~\ref{2e_harm_v_xc}(a)--(d)], and (c)--(d) the same quantities for the hard-wall confinement with radius $R$ [see Fig.~\ref{2e_harm_v_xc}(e)--(h)].
Panel (e) shows the same exchange-correlation potentials for 6 electrons with $(L,S)=(0,0)$ in harmonic confinement (see Fig.~\ref{6e_potentials}). Solid lines are the accurate results and dash lines the LSDA ones. Natural units determined by the external potential are used throughout.\label{v_xc_vs_n}}
\end{figure*}

In addition to plotting the model systems' exchange and correlation potentials as a function of the position we can analyze how they look when shown as a function of the local electron density and compare them in this representation with the LSDA. Deviations from the LSDA give a measure of the non-locality of the exchange and correlation effects. Figure~\ref{v_xc_vs_n} shows a comparison for our unpolarized systems with 2 or 6 electrons. Figures~\ref{v_xc_vs_n}(a)--(b) show the exchange-correlation and correlation potentials of the harmonically confined two-electron systems, both accurate and LSDA ones, of Fig.~\ref{2e_harm_v_xc}(b) and~(d) as a function of the local electron density, and Figs.~\ref{v_xc_vs_n}(b)--(d) those of the hard-wall systems of Figs.~\ref{2e_harm_v_xc}(f) and~(h). Figure~\ref{v_xc_vs_n}(e) shows the exchange-correlation potentials (see Fig.~\ref{6e_potentials}) of the 6-electron $(L,S)=(0,0)$ systems as a function of the local density. When looking at the total exchange-correlation potential, the accurate result and the LSDA one behave, in general, similarly. The vertical shift due to differing asymptotic behavior of the exchange potential is apparent. In the cases in which the accurate density is non-monotonous it becomes clear that the exchange-correlation potential cannot be expressed as a function of the local electron density but it is a non-local functional of the density. In the case of the 2-electron systems, we can identify the correlation potential, Figs.~\ref{v_xc_vs_n}(b) and~\ref{v_xc_vs_n}(d) as the main-source of the non-locality close to the density maxima. Other manifestations of the non-locality include the above-mentioned $-1/r$ asymptotics of the exchange potential absent from the LSDA one, see Figs.~\ref{2e_harm_v_xc}(c) and~(g), and the differing character of the exchange potential of the hard-wall systems, Fig.~\ref{2e_harm_v_xc}(g), see the discussion above. The behavior of the exchange-correlation potential as a function of the local density is similar between the 2 and 6-electron systems. Also in this comparison, the LSDA seems to work better for larger particle numbers. If Fig.~\ref{v_xc_vs_n} were plotted using atomic units, the LSDA potentials would collapse on top of each other. The universality seen in the accurate results of Figs~\ref{v_xc_vs_n}(a), \ref{v_xc_vs_n}(c), and \ref{v_xc_vs_n}(e) is an essentially non-local effect. It arises from the correct $-1/r$ asymptotics of the exchange potentials at the large $r$ (low local density) regime of the model systems.

\section{Conclusions}\label{conclusions}

In this work, we have extracted accurate exchange-correlation energies and potentials for inhomogeneous model electron systems in two dimensions starting from total energies and electron densities calculated by exact diagonalization. We have considered two electrons in a spin singlet in harmonic and circular hard-wall confinements and 4 electrons with $(L,S)=(0,1)$ and 6 electrons with $(L,S)=(0,0)$ in harmonic confinement. We have compared our results against results calculated by the local-spin-density approximation (LSDA) parametrization by Attaccalite \emph{et al.}~\cite{Attaccalite02} The LSDA appears to describe these systems relatively accurately.

Total energies of our model systems are curiously consistently overestimated by the LSDA. The exchange energy predicted by the LSDA is too high but this is partially compensated by the too low LSDA correlation energy. Considering the exchange and correlation potentials determining the accuracy of electron densities in density-functional calculations, the LSDA exchange potential is obviously lacking the exact $-1/r$ asymptotic tail because its lack of the self-interaction correction, a consequence of the locality of the LSDA. Nevertheless, the shape of the exchange-correlation potential is, apart from the two-electron systems in harmonic confinement, qualitatively correct, sometimes owing to error cancellation between the exchange and correlation components. In general, the LSDA as parametrized by Attaccalite \emph{et al.}~\cite{Attaccalite02} is quite accurate but it is clearly useful to go beyond the LSDA and further develop and test semi-local and non-local density and orbital functionals for electronic exchange and correlation in two-dimensions. Our results shown in this Article can act as benchmark data in creating, testing and parametrizing exchange and correlation functionals for two-dimensional electronic systems.

\begin{acknowledgments}
The authors acknowledge C.\ Webb and E.\ T\"ol\"o for their contributions to computer codes used in the study, and are thankful for discussions with E.\ R\"as\"anen and S.\ K\"ummel.
The authors also thank A.\ Uppstu for careful reading of the manuscript.
\end{acknowledgments}

\end{document}